\documentclass[conference]{IEEEtran}
\IEEEoverridecommandlockouts
\usepackage{cite}
\usepackage{amsmath,amssymb,amsfonts}
\usepackage{graphicx}
\usepackage{textcomp}
\usepackage{xcolor}
\def\BibTeX{{\rm B\kern-.05em{\sc i\kern-.025em b}\kern-.08em
    T\kern-.1667em\lower.7ex\hbox{E}\kern-.125emX}}
    
\usepackage{braket}
\usepackage{algorithm}
\usepackage{algpseudocode}
\usepackage{tikz}
\usepackage{hyperref}
\usetikzlibrary{quantikz}

\usepackage{flushend}

\begin{document}

\title{Transversal Injection: Using the Surface Code to Prepare Non-Pauli Eigenstates}

\newcommand{\linebreakand}{%
  \end{@IEEEauthorhalign}
  \hfill\mbox{}\par
  \mbox{}\hfill\begin{@IEEEauthorhalign}
}

% \author{\IEEEauthorblockN{Jason Gavriel}
% \IEEEauthorblockA{Centre for Quantum Computation \\ and Communication Technology.\\
% University of Technology Sydney\\
% Sydney, NSW, 2007, Australia\\
% jason@gavriel.au}
% \and
% \IEEEauthorblockN{Daniel Herr}
% \IEEEauthorblockA{d-fine GmbH, An der Hauptwache 7\\
% 60213, Frankfurt, Germany}
% \and
% \IEEEauthorblockN{Alexis Shaw}
% \IEEEauthorblockA{Centre for Quantum Computation \\and Communication Technology.\\
% University of Technology Sydney\\
% Sydney, NSW, 2007, Australia}
% \linebreakand
% \IEEEauthorblockN{Michael J. Bremner}
% \IEEEauthorblockA{Centre for Quantum Computation \\and Communication Technology.\\
% University of Technology Sydney\\
% Sydney, NSW, 2007, Australia}
% \and
% \IEEEauthorblockN{Alexandru Paler}
% \IEEEauthorblockA{\textit{Aalto University} \\
% Espoo, Finland}
% \and
% \IEEEauthorblockN{Simon J. Devitt}
% \IEEEauthorblockA{University of Technology Sydney\\
% Sydney, NSW, 2007, Australia}
% }

\author{
\IEEEauthorblockN{Jason Gavriel\IEEEauthorrefmark{1}\IEEEauthorrefmark{2}\IEEEauthorrefmark{5},
Daniel Herr\IEEEauthorrefmark{3},
Alexis Shaw\IEEEauthorrefmark{1}\IEEEauthorrefmark{2},
Michael J. Bremner\IEEEauthorrefmark{1}\IEEEauthorrefmark{2},
Alexandru Paler\IEEEauthorrefmark{4} and
Simon J. Devitt\IEEEauthorrefmark{2}}
\IEEEauthorblockA{\IEEEauthorrefmark{1}Centre for Quantum Computation and Communication Technology.}
\IEEEauthorblockA{\IEEEauthorrefmark{2}Centre for Quantum Software and Information \\ 
University of Technology Sydney.  Sydney, NSW, 2007, Australia.}
\IEEEauthorblockA{\IEEEauthorrefmark{3}d-fine GmbH, An der Hauptwache 7\\
 60213, Frankfurt, Germany}
\IEEEauthorblockA{\IEEEauthorrefmark{4}\textit{Aalto University}, Espoo, Finland}
\IEEEauthorblockA{\IEEEauthorrefmark{5}Email: jason@gavriel.au}}

% \author{Jason Gavriel}
% \email{jason@gavriel.au}
% \affiliation{Center for Quantum Software and Information, University of Technology Sydney.  Sydney, NSW, 2007, Australia.}
% \affiliation{Centre for Quantum Computation and Communication Technology.}
% \author{Daniel Herr}
% \affiliation{d-fine GmbH, An der Hauptwache 7, 60213, Frankfurt, Germany.}
% \author{Alexis Shaw}
% \affiliation{Center for Quantum Software and Information, University of Technology Sydney.  Sydney, NSW, 2007, Australia.}
% \affiliation{Centre for Quantum Computation and Communication Technology.}
% \author{Michael J. Bremner}
% \affiliation{Center for Quantum Software and Information, University of Technology Sydney.  Sydney, NSW, 2007, Australia.}
% \affiliation{Centre for Quantum Computation and Communication Technology.}
% \author{Alexandru Paler}
% \affiliation{Aalto University, 02150 Espoo, Finland.}
% \author{Simon J. Devitt}  
% \affiliation{Center for Quantum Software and Information, University of Technology Sydney.  Sydney, NSW, 2007, Australia.}

\maketitle

\begin{abstract}
The development of quantum computing systems for large scale algorithms requires targeted error rates unachievable through hardware advancements alone. Quantum Error Correction (QEC) allows us to use systems with a large number of physical qubits to form a fault tolerant system with a lower number of logical qubits and a favourable logical error rate. While some gates can be easily implemented in a QEC code transversally, there is no code that has a universal set of transversal gates. Transversal Injection is a new method of preparing logical non-Pauli eigenstates that can be used as resource states for quantum computation. State preparation can be done directly in the surface code and has the potential to prepare higher fidelity injected states. Compared to other techniques, transversal injection can reduce the resource burden for state distillation protocols. In this paper, the authors present the theory behind this new technique as well as an algorithm for calculating the resulting logical states prepared in the surface code.
\end{abstract}

\begin{IEEEkeywords}
quantum computing, surface code
\end{IEEEkeywords}

\section{Introduction}

As the development of quantum computing systems progresses over time it becomes increasingly difficult to achieve lower physical error rate. While this is tolerable for the NISQ regime \cite{SD-Preskill2018quantumcomputingin}, in order to realise large scale algorithms, some level of Quantum Error Correction (QEC) is necessary\cite{Devitt:2013aa,SD-Terhal:2015aa}. QEC allows us to utilise systems with a large number of physical qubits with error rates of $p \approx 10^{-3}$ as fault tolerant system with a lower number of logical qubits with a targeted error rate as low as $10^{-20}$ \cite{PhysRevX.8.041015,Matteo:2020aa}. The implementation of QEC naturally incurs an overhead in the number of physical qubits or in the computational time required \cite{Gidney2021}.

\subsection{Motivation}
The surface code \cite{SD-Fowler:2012aa} is a QEC code defined over a 2D nearest neighbour array of physical qubits. While there are plenty of competing QEC codes \cite{Hastings_2021,Breuckmann_2021}, the surface code is favourable in multiple hardware platforms \cite{Lekitsch:2017aa,SD-Jones:2012aa,Mukai:2020aa,SD-Hill:2015aa,Bombin21} and the most well studied. Some gates have a transversal realisation in the code which means the gate can be applied individually to all data qubits to perform a logical version of that gate. However, the Eastin and Knill no-go theorem \cite{East09} proves that any QEC code cannot simultaneously be error correcting and have a universal set of transversal gates. To work around this no-go, one can implement non-transversal gates through protocols such as Gauge-fixed code conversion \cite{SD-Bombin:2015aa} or a sequence of {\em state injection}, {\em magic state distillation} and {\em gate teleportation} \cite{Brav05,Bravyi:2012ul,Li_2015}. Even small changes in the fidelity of an injected state dramatically impacts the resources required for state distillation \cite{litinski19}, hence it has become a recent focus to look at better injection techniques \cite{singh2022high, gavriel2022transversal, gidney2023cleaner}.

\subsection{Contribution}

Transveral injection is a new method of preparing logical non-Pauli eigenstates that can be used as resource states for universal quantum computation. This technique can be generalised for all stabiliser codes, but we will focus on just the surface code for this paper. In standard surface code preparation, all data qubits are typically initialised into the $\ket{0}$ or $\ket{+}$ states. A series of stabilisers \cite{SD-Fowler:2012aa} are applied and measured and a logical Pauli state has been successfully prepared. 

Transversal injection is a relative simple protocol that modifies the initialisation step of a surface code. A uniform, single qubit rotation is performed on each data qubit before the first round of stabilisers are applied. The stabiliser measurements will collapse our system into some non-Pauli logically encoded state which can now be uitilised in distillation and teleportation protocols to achieve universality int he surface code.

The state produced is both {\em probabilistic} and {\em heralded}. It can be derived from the definition of the code, the angle of the initial rotation and the string of measurement outcomes. This string of initialise stabiliser measurements made when initialising the code, which we will call a {\em stabiliser trajectory}, determines what particular state we have prepared. In this paper, we detail an algorithm for calculating the resulting logical states of the transversal injection protocol and discuss the implications for fault-tolerant quantum computing.

Transversal injection prepares a single, logically encoded non-Pauli state out of a set of possible states. The state produced is probabilistic and not known ahead of time although the set can be computed ahead of time for low code distances. For code distances where the number of possible states becomes exponentially large, an algorithm is provided to calculate the resultant state on the fly. 

\section{Method: Transversal Injection Protocol}

Preparation of resource states to create a universal gate-set is typically a costly part of quantum computation \cite{Brav05,Litinski2019gameofsurfacecodes,litinski19}. Numerical simulations have demonstrated that the fidelity of a state produce by transversal injection has a linear relationship with the physical error rate \cite{gavriel2022transversal} which is comparable to what is capable with Li state injection \cite{Li_2015}. Post-selection strategies have the potential to further reduce the logical error rate and any reduction at this point of the compilation pipeline can have dramatic impacts on the resources requirements.

Herein, we describe how a non-Pauli logical state can be encoded on a surface code using only transversal operations. 

\subsection{Initialisation}
When preparing the surface code, the data qubits are usually initialised into the $|0\rangle_L$ or $|+\rangle_L$ states as these states are eigenstates of the code. Let's instead consider the case where all the data qubits are initially placed into the states
\begin{equation}
|\chi\rangle = \alpha|0\rangle + \beta|1\rangle
\end{equation}
i.e. we first perform a transversal rotation on all data qubits to the state $|\chi\rangle$ before we follow the same procedure to encode a $|0\rangle_L$ or $|+\rangle_L$ state.

When considering the system of $N$ data qubits that have been uniformly transformed, the values of each eigenstate are now naturally related to the Hamming weight of its integer value and the chosen transversal operation. 
\begin{equation}
\ket{\chi}^{\otimes N} = \sum^{2^N-1}_{n = 0} \alpha^{N-\hat{H}(n)} \beta^{\hat{H}(n)} \ket{n}
\label{eq:ti_init}
\end{equation}

Where $n$ is the integer representation of the system's eigenvalues and $\hat{H}(n)$ is the Hamming weight, or number of bits set to 1 in its binary representation. 

\subsection{Syndrome Extraction}
To encode our logical state, we need to measure the stabilisers of the surface code \cite{SD-Fowler:2012aa}. The ancillary qubits (coloured grey in Fig.~\ref{fig:d2_unrot_layout}, representing the unrotated surface code $d=2$) are measured to observe the party across the $X$-type vertex stabilisers and the $Z$-type plaquette stabilisers. 

\begin{figure}[!t]
    \centering
	\includegraphics[width=3.5cm]{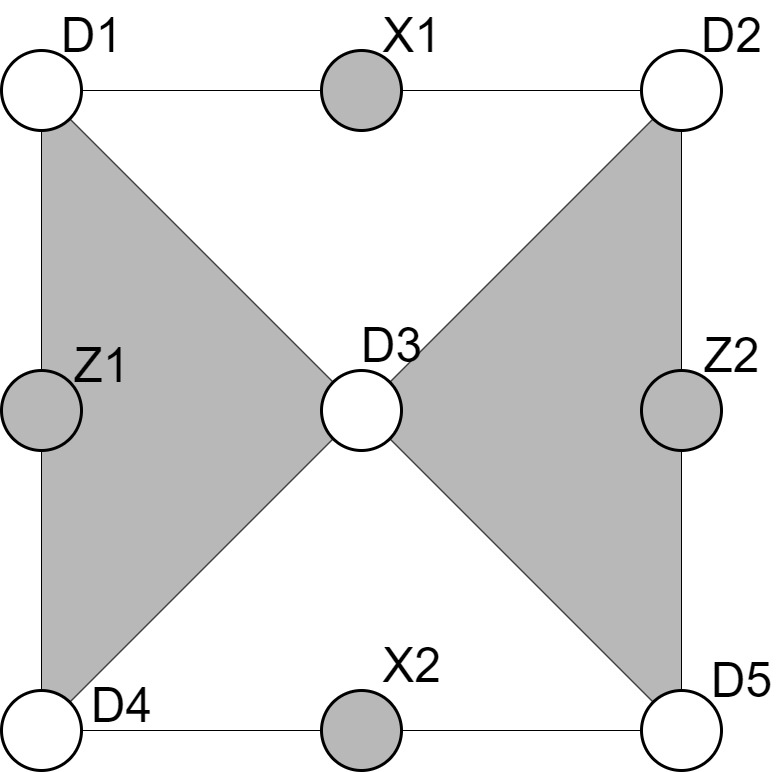}
	\caption{Qubit layout of the unrotated surface code at distance 2}
	\label{fig:d2_unrot_layout}
\end{figure}

The set of measurements for each stabiliser $\{X_{v_i}, Z_{p_i}\}$ defines a {\em stabiliser trajectory}. This set is denoted by the string $\{X_1,...,X_{(N-1)/2}, Z_1,...,Z_{(N-1)/2}\} \in \{0,1\}$ of eigenvalues, that are measured for the $N-1$ $X$-type vertex and $Z$-type plaquette stabilisers. This definition will be used throughout this paper to reference a set of measurement outcomes that together determine a resultant logical state. In the absence of physical errors, the $X$ projections are probabilistic and the $Z$ projections are deterministic when encoding a $|0\rangle_L$ or vice versa for the $|+\rangle_L$ state. In transversal injection the outcomes of {\em all} measurements are probabilistic and will select out only certain eigenvalues that are consistent with the previously observed measurements.

The first step in encoding is to measure all the $X$-type stabilisers, $X_{v_i}$. Our initial state is projected into eigenstates of the $X$-stabilisers with the eigenvalue for each operator determined by the qubit measurements of each stabiliser circuit. The measurement of these stabilisers will result in an eigenvalue sequence for the $X$-type stabilisers and cause perturbations in the superposition of the $Z$-type stabilisers until they too are measured. The second step is to project into eigenstates of the $Z$-stabilisers in the same manner.

When we initialise into the $|0\rangle_L$ state, we would now already have a logically encoded state (as all the $Z$-type stabilisers {\em including} $Z_L$ should automatically be satisfied), however these stabilisers are still measured anyway for a fault-tolerant initialisation. In the case of transversal injection $Z_L$ is now in some superposition of states, determined by the stabilisers that either commute or anti-commute — our stabiliser trajectory.

The logical state of the block is $z_L \in \{0,1\}$ where $0$ indicated a $+1$ eigenstate of the logical Z operator ($\ket{0}_L$) and $1$ as the $-1$ eigenstate ($\ket{1}_L$). The chain operator is also determined as,
\begin{equation}
z_{\text{L}} = \left(\sum_{i \in \text{left/right}}z_i\right) \mod 2. 
\end{equation}
where left/right is the set of qubits of any chosen chain that runs from the left boundary of the lattice to the right (defining the logical $Z$ operator). If we physically measured each data qubit in the chain, the parity of these measurements would give us $z_L$.

When all qubits are initialised into $\ket{\chi}^{\otimes N}$ before encoding, we now are not simply in $\pm 1$ eigenstates of $Z_{p_i}$. Instead we are in a superposition of the $0$ and $1$ eigenstates. Using the syndrome measurements, we know how the stabiliser sets transform from the original diagonal $Z$ form to the stabilisers and logical operators for the encoded planar surface code. To calculate the resultant logical state, we now treat each term individually and follow them through the encoding procedure. 

\subsection{Resultant state}
As we are no longer in eigenstates of the logical $Z$ operator of the surface code, we will be left with some non-trivial logically encoded state. The probabilistic nature of the $X$ and $Z$-type stabiliser measurements means that the resultant encoded state from transversal injection is also probabilistic and cannot be known ahead of time. The fact that we follow only one of the $2^{(N-1)}$ possible stabiliser trajectories when we initialise the code means we select out only the terms that are consistent with the eigenvalues we measure. For example, in the d=2 example in Fig.~\ref{fig:d2_unrot_layout}, if $Z_{p_1} = 0$, for stabiliser $Z_{p_1} = Z_{D1}Z_{D3}Z_{D4}$, then we can only keep the terms where $Z_{p_1} = (z_{D1} + z_{D3} + z_{D4}) \mod 2 = 0$.  Any term where $Z_{p_1} = 1$ will be inconsistent with the measurement projection that actually occurred and will be simply projected out of the resultant state.

What we are left with is a superposition of {\em only} the eigenstates that are consistent with the $X$- and $Z$-stabiliser trajectories that are observed when initialising the state.  By definition - as we have now stabilised our data qubits with respect to {\em all} stabilisers of the planar surface code - we are going to be left with {\em some} superposition of the eigenstates of the logic operator, $Z_L$, i.e. some new encoded state, $\ket{\Lambda}_i = \alpha_L|0\rangle_L + \beta_L|1\rangle_L$.  
Note that the resultant {\em logical} state does not, in general, match the transversal state, $|\chi\rangle$, that was used to initialise each of the physical qubits in the encoded block.

To calculate the resultant logical state, we examine the last entry of the eigenvalue vector which is the logical observable, $L_i = z_{\text{chain}}$, formed from the eigenvalues of the individual data qubits that form a connected left/right chain through the planar surface code. The amplitude of the resulting $|0\rangle_L$ state, $\alpha_L$, will be the sum of all the terms where $L_i = 0$, while the amplitude of the $|1\rangle_L$, $\beta_L$ will be the sum of all the terms where $L_i=1$. After re-normalising the wave-function, we can now have an analytical form of the resultant encoded state as a function of $\alpha$, $\beta$ and $N$.  

\section{Results: Algorithm for calculating a logical state}

Algorithm~\ref{alg2} allows us to determine the resultant state without a full state simulation. We begin by determining which eigenstates commute with the $Z$-stabiliser measurement outcomes (lines 3-19). For non-trivial $X$-trajectories, we now project onto the target $X$-stabilisers which forms the bulk of the computation.

\label{link2}
\begin{algorithm}[!t] 
\caption{Calculating a specific logical state.}
\label{alg2}
\begin{algorithmic}[1]
\Require{$\hat{X} = X_{1} \dots X_{(N-1)/2}$, $X$-stabilisers.} 
\Require{$\hat{Z} = Z_{1} \dots Z_{(N-1)/2}$, $Z$-stabilisers.} 
\Require{$L$, a chain operator for the logical $Z$-state.}
\Require{$\{\alpha$,$\beta\}$, coefficients of transversal physical states}
\Require{Number of data qubits of a distance $d$ planar surface code, un-rotated code, $N=d^2 + (d-1)^2$}
\Require{$\langle x_i \rangle, \langle z_i \rangle$ eigenvalue bit strings corresponding to stabiliser measurements}

\Statex
\State {$\hat{m} = \mathbf\{\}$ \# List of set stabiliser measurements $(N-1)/2$}
\State {$\hat{v} = [\{0\}_N]$} \#Eigenvalue memory, initially just $\{0\}_N$

\For{$i$ in $(N-1)/2$}
    \State {$\hat{v_t} = []$} \#Loop eigenvalue storage
    \For{$v$ in $\hat{v}$}
        \State {$t = z_i$}
        \For{$m$ in $\hat{m}$}
            \State{ $t = t \oplus (m_i)$ }
        \EndFor
        \State{$\hat{u} = \neg \hat{m}$}    \# ignoring $Z_i$ not adjacent to $i$th qubit
        \State $\hat{c} = \sum_{n = 0}\binom{\hat{u}}{2n+t}$ \# combinations for parity $t$
        \State{Append $Z_i$ to $\hat{m}$}
        \For{$c$ in $\hat{c}$}
            \State{$\hat{e} = \hat{z} \lor c$}
            \State{Append $e$ to $\hat{v_t}$}
        \EndFor
        \State{$\hat{v} = \hat{v_t}$}
    \EndFor
\EndFor

\State $\alpha_L =$ $\beta_L = 0$
\For{$v$ in $\hat{v}$}
    \State{$j = bitwise\_sum(\hat{z})$}
    \State {$\hat{k} = \hat{v} \times \hat{x}$}
    \For{$k$ in $\hat{k}$}
        \State $l = k \times B$ \# Parity check with logical operator
    	\State {\bf if} $l = 0$, $\quad \alpha_L$ = $\alpha_L + \alpha^j \beta^{N-j}$.
    	\State {\bf if} $l = 1$, $\quad \beta_L$ = $\beta_L + \alpha^j \beta^{N-j}$.
    \EndFor
\EndFor
\State \Return {$\{\alpha_L$, $\beta_L\}/\sqrt{|\alpha_L|^2+|\beta_L|^2}$}
\end{algorithmic}
\end{algorithm}

For example, consider a distance 2 unrotated surface code (Fig.~\ref{fig:d2_unrot_layout}) with 5 data qubits and a target $\hat{0001}$. The two $Z$-stabilisers of the distance two surface code form the matrix below.
\begin{equation}
M = \begin{pmatrix}
D_1 & D_2 & D_3 & D_4 & D_5 \\
1 & 0 & 1 & 1 & 0 \\
0 & 1 & 1 & 0 & 1 
\end{pmatrix}
\label{eq:stabmatrix}
\end{equation}
If we create a representation of the $Z$-stabilisers where we store the index of elements equal to one, we have a directory of which data qubits each $Z$-stabiliser impacts. In this example we use 0 as the first index.
\begin{equation}
M_{aux} = \begin{pmatrix}
0 & 2 & 3 \\
1 & 2 & 4
\end{pmatrix}
\end{equation}

For the first $Z$-stabiliser we look up the first bit of $\hat{01}$, which is $0$. Consider all combinations of $\{0, 2, 3\}$ that are of length $\{0, 2\}$ (or $\{1, 3\}$ if it were equal to $1$). We now iterate through each of $\{(\_), (0, 2), (0, 3), (2, 3)\}$ and set the bits of our initial trajectory $\{0\}_N$.

\begin{equation}
\hat{v_t} = \begin{pmatrix}
0 & 0 & 0 & 0 & 0 \\
1 & 0 & 1 & 0 & 0 \\
1 & 0 & 0 & 1 & 0 \\
0 & 0 & 1 & 1 & 0 
\end{pmatrix}
\end{equation}

For each subsequent $Z$-stabiliser, we set $t$ equal to the $i$th bit of $\hat{01}$. Depending which values of $\{1, 2, 4\}$ have been set already, we create two subsets for the seen and unseen indices (i.e. $\{2\}$ and $\{1, 4\}$). For each result from the previous loop, we {\bf xor} $t$ with the value of each bit at the seen indices. As in the first case, we iterate through all combinations of unseen bits $\{1, 4\}$ that are odd or even lengths depending on $t$. Again we set the bits of the result at the indices — for each new combination, for each result from the previous loop. In our example below, $t$ will alternate between 0 and 1 as $z_2$ is our unseen and as $t$ is initialised as $1$ for the second bit, results will be flipped.

{\footnotesize

    \begin{equation}
    \begin{bmatrix} t & z_0 & z_1 & z_2 & z_3 & z_4 & $combinations$ \\
    1 & 0 & 0 & 0 & 0 & 0 & (Z_{p_1}), (Z_{p_4})\\
    0 & 1 & 0 & 1 & 0 & 0 & (\_), (Z_{p_1}, Z_{p_4})\\
    1 & 1 & 0 & 0 & 1 & 0 & (Z_{p_1}), (Z_{p_4})\\
    0 & 0 & 0 & 1 & 1 & 0 & (\_), (Z_{p_1}, Z_{p_4})
    \end{bmatrix}
    \end{equation}
    
    \begin{equation}
    Results = \begin{bmatrix} z_0 & z_1 & z_2 & z_3 & z_4 & $j$ \\
    0 & 0 & 0 & 0 & 1 & 1 \\
    0 & 1 & 0 & 0 & 0 & 1 \\
    1 & 0 & 1 & 0 & 0 & 2 \\
    1 & 1 & 1 & 0 & 1 & 4 \\
    1 & 0 & 0 & 1 & 1 & 3 \\
    1 & 1 & 0 & 1 & 0 & 3 \\
    0 & 0 & 1 & 1 & 0 & 2 \\
    0 & 1 & 1 & 1 & 1 & 4 
    \end{bmatrix}
    \end{equation}
}

For each row, we determine Hamming weight $j$ from the count of ones in $\hat{z}$. Note $j$ will remain the same during the next step even though the Hamming weight will change. 

It should be noted that after any loop in the algorithm, the intermediate results can be processed in parallel with no penalty in complexity aside from the negligible overhead of distributing the work. Using GPU acceleration, we have demonstrated sampling from the set of all trajectories with trivial $X$-syndromes for stabiliser codes with 113 data qubits ($d = 8$, un-rotated).

If we are now to examine a non-trivial $X$-syndrome, an additional step is needed. For example, let's examine a target trajectory of $\hat{1001}$ where the first two bits are our $X$-stabiliser measurement outcomes. We now project the results from our last step over these values as below.

{\footnotesize
\begin{equation}
\begin{bmatrix}[c|cccc|c|c] \hat{z} & II & X_1 & X_2 & X_1 X_2 & $j$ & $k$\\
\hat{00001} & \hat{00001} & - \hat{11101} & \hat{00110} & - \hat{11010} & 1 & 0\\
\hat{01000} & \hat{01000} & - \hat{10100} & \hat{01111} & - \hat{10011} & 1 & 1\\
\hat{10100} & \hat{10100} & - \hat{01000} & \hat{10011} & - \hat{01111} & 2 & 1\\
\hat{11101} & \hat{11101} & - \hat{00001} & \hat{11010} & - \hat{00110} & 4 & 0\\
\hat{10011} & \hat{10011} & - \hat{01111} & \hat{10100} & - \hat{01000} & 3 & 1\\
\hat{11010} & \hat{11010} & - \hat{00110} & \hat{11101} & - \hat{00001} & 3 & 0\\
\hat{00110} & \hat{00110} & - \hat{11010} & \hat{00001} & - \hat{11101} & 2 & 0\\
\hat{01111} & \hat{01111} & - \hat{10011} & \hat{01000} & - \hat{10100} & 4 & 1
\end{bmatrix}
\end{equation}
}

We determine $k$ from the parity of the first $d$ bits (our logical observable, row 5 of Eq. \ref{eq:stabmatrix}) and these values can be used to construct the analytical equations (prior to re-normalisation) for our target. Each term contributes $\alpha^j \beta^{N-j}$ to the corresponding logical state, and we can collect them as below.

{\footnotesize
\begin{equation}
\begin{aligned}
\begin{pmatrix} \ket{00001} \\ \ket{00110} \\ \ket{11101} \\ \ket{11010}  \\ \ket{01000}  \\ \ket{01111}  \\ \ket{10100}  \\ \ket{10011}\end{pmatrix} = \begin{pmatrix} 
2\alpha^4\beta + 2\alpha^3\beta^2  - 2\alpha^2\beta^3 - 2\alpha\beta^4 \\ 
2\alpha^4\beta + 2\alpha^3\beta^2  - 2\alpha^2\beta^3 - 2\alpha\beta^4 \\ 
-2\alpha^4\beta - 2\alpha^3\beta^2  + 2\alpha^2\beta^3 + 2\alpha\beta^4 \\ 
-2\alpha^4\beta - 2\alpha^3\beta^2  + 2\alpha^2\beta^3 + 2\alpha\beta^4 \\ 
2\alpha^4\beta - 2\alpha^3\beta^2  - 2\alpha^2\beta^3 + 2\alpha\beta^4 \\ 
2\alpha^4\beta - 2\alpha^3\beta^2  - 2\alpha^2\beta^3 + 2\alpha\beta^4 \\ 
- 2\alpha^4\beta + 2\alpha^3\beta^2  + 2\alpha^2\beta^3 - 2\alpha\beta^4 \\ 
- 2\alpha^4\beta + 2\alpha^3\beta^2  + 2\alpha^2\beta^3 - 2\alpha\beta^4 \end{pmatrix} \\
\end{aligned}
\label{eq:4eqn}
\end{equation}
}

Due to the Pauli frame set by our stabiliser measurements, what we consider our logical state is a mix of the following terms.

\begin{equation}
\begin{multlined}
\ket{\Lambda} = \begin{pmatrix} \alpha_L \\ \beta_L \end{pmatrix} = \begin{pmatrix} 
\ket{00001} + \ket{00110} - \ket{11101} - \ket{11010}  \\ - \ket{01000}  - \ket{01111}  + \ket{10100}  + \ket{10011} \end{pmatrix}
\\
\ket{\Lambda} = \begin{pmatrix} 
\alpha^4\beta + \alpha^3\beta^2  - \alpha^2\beta^3 - \alpha\beta^4 \\  - \alpha^4\beta + \alpha^3\beta^2  + \alpha^2\beta^3 - \alpha\beta^4 \end{pmatrix} \\
\end{multlined} 
\label{eq:5eqn}
\end{equation}

It should be noted that any non-trivial syndrome will introduce bit and/or phase flips to the logical code-words. As is typical of a stabiliser code, corrections need to be applied to move back into an even superposition of $\ket{0}_L$ and $\ket{1}_L$, or stored within classical memory (i.e. tracking the Pauli frame).

Algorithm~\ref{alg2} will allow for the calculation of any transversally realisable encoded state for an arbitrarily large distance code, $d$: it intuitively follows the logic of the stabiliser execution.

\subsection{Algorithm Complexity and Just-In-Time Operation}
To simulate all possible resultant states, one must perform a full state simulation and track how the system evolves from the initial state to the encoded state as the stabilisers commute or anti-commute during runtime. This requires enough memory to store the full state of the logical and ancilla qubits which quickly becomes difficult to simulate classically. This algorithm instead tracks only the terms relevant to a single target trajectory, and we can now compute the logical state of this trajectory in a more efficient manner. 

The algorithm scales exponentially with the number of $Z-$stabilisers, rather than with the number of physical qubits required for the full state simulation. This improvement is most noticeable in the memory footprint which can now be dealt with comfortably as we only store $e^{(N-1)/2}$ terms in memory. Instead of looping over each term, this computation can be done in parallel and using GPU acceleration we have demonstrated this algorithm on trajectories with trivial $X$-measurements on systems with over 100 data qubits. When considering non-trivial $X$ measurements the complexity is $\mathcal{O}(e^{(N-1)})$ as we must iterate over the $e^{(N-1)/2}$ $X$-stabilizers. These additional steps for calculating for non-trivial $X$-stabiliser measurements are costly and this classical algorithm is only feasible for code distances $<$ 6.

This improvement from a full state simulation opens up the opportunity for a just-in-time strategy where output states from transversal injection can be calculated classically and then used as part of a broader compilation strategy. To realise this strategy, a more efficient algorithm will be needed for larger distances required for large-scale, error-corrected computation.

\begin{figure}[!t]
    \centering
	\includegraphics[width=0.7\columnwidth]{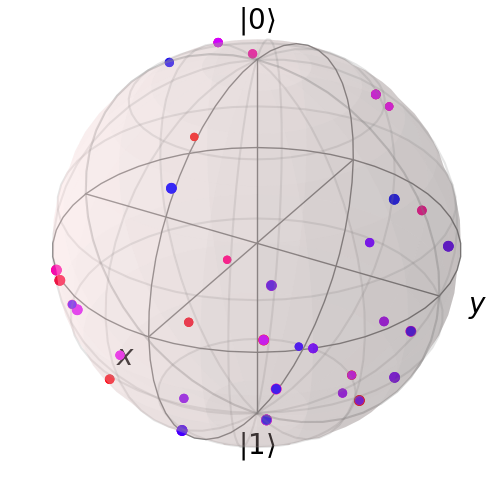}
	\caption{The 64 logical, single qubit, states (only trivial $X$-stabiliser measurements), resultant from transversal injection on a distance $d=3$ unrotated code from the table above. Generated using QuTiP~\cite{johansson_qutip_2013}.}
	\label{fig:d3}
\end{figure}

\subsection{Practical Implications}

Transversal injection provides us with a large family of possible states, the number of which scaling exponentially as a function of data qubits. The probability of a particular state being prepared becomes exponentially unlikely the more states we have to choose from and there is no way of controlling the state that comes out. What can be controlled is the distribution of possible states through choice of stabiliser code and initial rotation. There are likely redundancies in the distributions and compilation techniques that can be utilised to efficiently compile error-corrected circuits.

In the figure below, we have plotted the distribution of resultant states across the Bloch sphere for a range of initial parameters. There are clear structures that form depending on the parameters of the initial rotation including great circles, arcs with different radii, clusters around poles and dense packing of the entire sphere. Certain distributions may be advantageous for certain compilation strategies or for physical systems that have error biases in certain directions. Alternatively, having a dense covering of the Bloch sphere might be optimal in a compilation strategy that benefits from a large set of distinct states. 

\begin{center}
    \includegraphics[width=1\columnwidth]{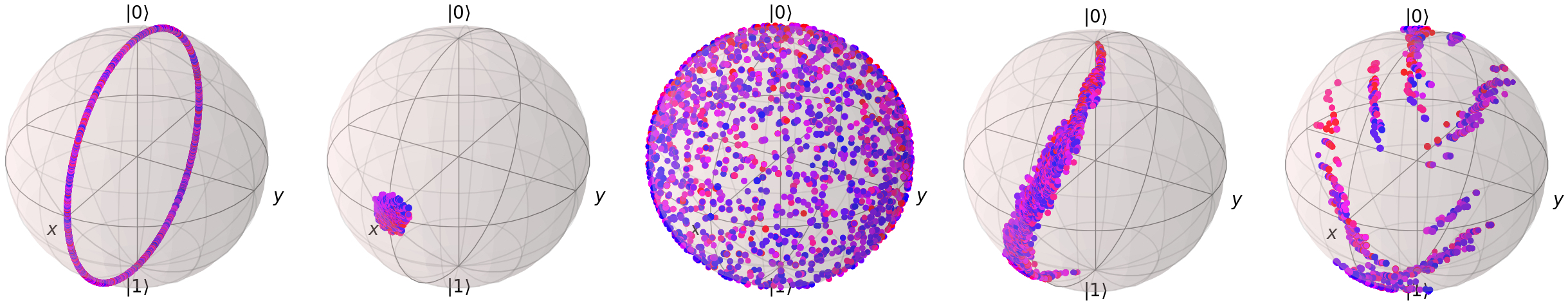}
\end{center}   

One example of a strategy is simply a series of random walks around SU(2) \cite{Brav05}. An alternative method is to solve the analytical equations given by the above algorithm and to post-select for trajectories that provide the desired state. 
Some numerical simulations have been done under a simple symmetric Pauli error model \cite{gavriel2022transversal} and demonstrate that a logical error rate {\em lower} than the physical error rate is achievable using this protocol. Magic states prepared in this fashion can then be distilled to an arbitrary accuracy, although further research into distilling arbitrary states would be useful in conjunction with this new technique. 

\section{Conclusion}

Transversal injection is a new technique for preparing non-Pauli eigenstates directly into any stabiliser code. This allows universal quantum computation without a separate magic state preparation protocol and provides a logical gate-set that isn't constrained to simply Clifford + $T$.

While some numerical simulations have been done, a direct resource comparison requires a mature compilation technique and further research into resource estimation. Even though the protocol is probabilistic, the advantages of using a more direct state preparation technique can outweigh this disadvantage depending on the target hardware architecture and may lead to a potential reduction in qubit requirements for any large-scale algorithm.  

While a direct resource comparison to a compiled algorithm using magic state distillation, such as Shor-2048 \cite{Gidney2021} will require a systematic solution to compiling arbitrary single qubit logical rotations, there is a potential for reduced qubit requirements for any large-scale algorithm by utilising this new technique.

\section*{Acknowledgements}
Thank you to Austin Fowler for early feedback on the findings of this paper. The views, opinions and/or findings expressed are those of the authors and should not be interpreted as representing the official views or policies of the Department of Defense or the U.S. Government. This research was developed in part with funding from the Defense Advanced Research Projects Agency [under the Quantum Benchmark- ing (QB) program under award no. HR00112230007 and HR001121S0026 contracts]. MJB acknowledges the support of Google. MJB, JG, and AS, were supported by the ARC Centre of Excellence for Quantum Computation and Communication Technology (CQC2T), project number CE170100012. AS was als supported by the Sydney Quantum Academy.

\bibliographystyle{unsrt}
\bibliography{bib2}

\end{document}